\documentclass[cits]{PoS}
\usepackage{amsmath,amssymb,mathbbol}
\graphicspath{{./Figures/}}

\newcommand{\preprint}{
  \begin{picture}(0,0)
    \put(0,110){{\rm\normalsize ADP-07-09/T649, HU-EP-07/48}}
  \end{picture}}

\title{\preprint%
Running {\huge$\alpha_s$} from Landau-gauge gluon and ghost correlations}

\ShortTitle{$\alpha_s$ from gluon and ghost correlations}

\author{\speaker{A.~Sternbeck}, K.~Maltman\thanks{And Department of Mathematics
    and Statistics, York Univ., Toronto, ON, M3J 1P3, Canada.}, 
    L.~von Smekal and A.~G.~Williams\\  
        Centre for the Subatomic Structure of Matter
        (CSSM), Department of Physics,\\
        University of Adelaide, SA 5005, Australia\\
        E-mail: \email{andre.sternbeck@adelaide.edu.au}}

\author{E.-M.~Ilgenfritz and M.~M{\"u}ller-Preussker\\
        Humboldt-Universit\"at zu Berlin, Institut f\"ur Physik,
        12489 Berlin, Germany}

\abstract{We estimate the running coupling constant of the strong
  interactions within the nonperturbative framework of lattice QCD in
  Landau gauge. Our calculation is based on the ghost-gluon vertex
  which in the particular case of Landau gauge allows for a definition
  of $\alpha_s$ in a MOM scheme solely in terms of the gluon and ghost
  dressing functions. As a first step we investigate the zero and
  two-flavour case and report here on preliminary results.}

\FullConference{The XXV International Symposium on Lattice Field Theory\\
		 July 30 -- August 4, 2007\\
		 Regensburg, Germany}

\newcommand{\Eq}[1]{Eq.~(\ref{#1})}
\newcommand{\Fig}[1]{Fig.~\ref{#1}}
\newcommand{\Tab}[1]{Tab.~\ref{#1}}

\newcommand{\MSbar}{\mathrm{\overline{MS}}}
\newcommand{\MOM}{\mathrm{MOM}}
\newcommand{\alphaMS}{\alpha_s^{\MSbar}}
\newcommand{\LambdaMS}{\Lambda^{\MSbar}}
\newcommand{\alphaMOM}{\alpha_s^{\MOM}}
\newcommand{\Tr}{\operatorname{Tr}}                 
\newcommand{\identity}{\mathbb{1}}                  
\newcommand{\RS}{\mathtt{S}}

\begin{document}

\section{Introduction}

The renormalised QCD coupling $\alpha_s=g^2/(4\pi)$ is one of the
fundamental parameters of the Standard Model of particle physics and
can not only be determined from high-energy experiments but also 
estimated directly in lattice QCD simulations (see, e.g.,
\cite{Prosperi:2006hx} for a recent review).

The actual values of $\alpha_s$ depend on both the renormalisation 
scheme (including the number of active flavours) and the scale
considered. Given a renormalisation scheme $\RS$ the  
dependence of $\alpha^\RS_s$ on the scale $\mu$ is controlled by the
renormalisation group through
\begin{equation}
  \mu^2\frac{d}{d\mu^2}\frac{\alpha^\RS_s(\mu^2)}{\pi} =
  \beta^\RS\left(\alpha^\RS_s \right)
  \; \sim \;  - \sum_{i\ge0}\beta^\RS_i\left(\frac{\alpha^\RS_s}{\pi}
  \right)^{i+2},
  \label{eq:RGE}
\end{equation}
where $\beta^\RS$ is the beta function defined in that scheme. 
Its asymptotic expansion is known up to four loops in the 
$\MSbar$ scheme \cite{vanRitbergen:1997va} and up to
three loops in various $\MOM$ schemes \cite{Chetyrkin:2000fd}. The 
first two coefficients are renormalisation-scheme independent and can
be looked up, e.g., in \cite{Yao:2006px} and references therein. 
The number of physically admissible renormalisation schemes (RSs) is
unlimited, and so is the number of running couplings. The
$\MSbar$ scheme, with the underlying use of dimensional
regularisation, is currently the most widely used RS in the analysis
of high-energy experimental data. Such experiments are usually
performed at different scales $\mu_i$, but through \Eq{eq:RGE} the
different values of $\alphaMS(\mu_i)$ are related to each other.%
\footnote{In the literature it is common practice to evolve data to $\alphaMS$
at the $Z$-Boson mass (see, e.g., \cite[Fig.23]{Davier:2005xq} for a nice
illustration.) A recent compilation of $\alphaMS(M_{Z})$ values has been
given at the ICHEP conference resulting in the world average
$\alphaMS(M_{Z})=0.1175\pm0.0011$ \cite{Kluth:2006vf}.}
In order to keep track of different initial parametrisations
$(\mu_i,\alphaMS(\mu_i))$ a scale-invariant parameter $\LambdaMS$ 
is introduced. This can be done for any RS. Once the $\Lambda$ parameter is
known in one RS, a one-loop calculation is sufficient to obtain it in any 
other scheme (see, e.g., \cite{Celmaster:1979km}). 

Various nonperturbative studies in the past, in particular in lattice QCD,
have provided estimates for $\alphaMS(M_{Z})$ or $\LambdaMS$ for
different numbers of flavour (see, e.g., 
\cite{Mason:2005zx,Gockeler:2005rv,Capitani:1998mq,DellaMorte:2004bc}). 
There, different RSs, and thus different nonperturbative definitions of
$\alpha_s$, have been used and the results roughly agree with what has
been found in experiment. Our objective here is to show that, in future,
good results for $\alphaMS(M_{Z})$ or $\LambdaMS$ may be expected 
using lattice QCD in Landau gauge.

The definition of the running coupling we employ here is a
nonperturbative one that has been first presented in the
context of introducing a solvable systematic truncation scheme for the 
Dyson-Schwinger equations of Euclidean QCD in Landau gauge
\cite{vonSmekal:1997isvonSmekal:1997vx}. This running coupling, which
we call $\alphaMOM$ in what follows, is defined in a $\MOM$ scheme and
has its seeds in the ghost-gluon vertex in Landau gauge. 
To be specific, we use \cite{vonSmekal:1997isvonSmekal:1997vx}
\begin{equation}
  \label{eq:running_coup}
  \alphaMOM(q^2) := \alphaMOM(\mu^2)\, Z(q^2,\mu^2)\, J^2(q^2,\mu^2)
\end{equation}
which defines a nonperturbative running coupling that enters directly
into the DSEs of QCD \cite{Alkofer:2000wg}. $Z$ and
$J$ are the renormalised dressing functions of the gluon and ghost
propagators, 
\begin{equation}
  \label{eq:gl_gh_tensor}
 D^{ab}_{\mu\nu}(q^2,\mu^2) =
 \delta^{ab}\left(\delta_{\mu\nu} - \frac{q_{\mu}q_{\nu}}{q^2}
 \right)\frac{Z(q^2,\mu^2)}{q^2}\quad\textrm{and}\quad G^{ab}(q^2,\mu^2) =
 -\delta^{ab}\frac{J(q^2,\mu^2)}{q^2}\,.
\end{equation}
In a lattice regularised theory, the
corresponding bare dressing functions $Z_L$ and $J_L$ are related to
$Z$ and $J$, renormalised at some sufficiently large $q^2=\mu^2$, by 
\begin{equation}
  \label{eq:ren_dressing_to_bare}
   Z(q^2,\mu^2) = Z^{-1}_3(\mu^2,a^2)\; Z_L(a^2,q^2)
   \quad\text{and}\quad
   J(q^2,\mu^2) = \widetilde{Z}^{-1}_3(\mu^2,a^2)\; J_L(a^2,q^2)
\end{equation}
where $Z_3$ and $\widetilde{Z}_3$ are the respective renormalisation
constants and $a$ is the lattice spacing. When considering $1/a$ as our
lattice UV cutoff, i.e., for $a\to0$, then %
\footnote{Note that, in Landau gauge, as shown long 
  ago \cite{Taylor:1971ffMarciano:1977su}, the ghost-gluon vertex 
  is regular and finite to any order in perturbation theory. Its
  renormalisation constant can thus be set to 
  $\widetilde{Z}_1=1$. Numerical evidence that this is also valid
  nonperturbatively has been provided in various investigations (see, e.g.,
  \cite{Alkofer:2004it,Cucchieri:2004sqIlgenfritz:2006he,Sternbeck:2006rd}).}
\begin{equation}
  \frac{g^2(a)}{4\pi}\;
  \frac{Z_3(\mu^2,a^2)\;\widetilde{Z}^2_3(\mu^2,a^2)}{\widetilde{Z}_1(\mu^2,a^2)}
  \quad\stackrel{a\to0}{\longrightarrow}\quad\alphaMOM(\mu^2)\;.
\end{equation}
With discretisation errors of $O(a^2)$ we can thus write
\begin{equation}
  \label{eq:running_coup_latt}
  \alphaMOM(q^2) = \frac{g^2(a)}{4\pi}\, Z_L(q^2,a^2)\,
  J^2_L(q^2,a^2) + O(a^2)\,,
\end{equation}
where $g^2(a)$ is the bare coupling at the lattice cutoff scale
$1/a$. It is this form of the running coupling which we will use below.

\section{Details of the numerical simulation}

The preliminary results described here were obtained on both zero and
two-flavour $SU(3)$ gauge field configurations. The quenched
configurations were thermalised using the standard Wilson gauge action
at several values of $\beta \equiv 6/g^2(a)$. We applied update cycles
each consisting of one heatbath and four micro-canonical 
over-relaxation steps. The unquenched gauge field configurations were
provided to us by the QCDSF collaboration. They used the same gauge action
but supplemented it by $N_f=2$ clover-improved Wilson fermions at various
values of the hopping-parameter~$\kappa$. For details on the choice of
$\beta$- and $\kappa$-values we refer to \Tab{tab:stat}. All gauge
configurations were fixed to Landau gauge with an iterative
Fourier-accelerated gauge-fixing
algorithm~\cite{Davies:1987vs}.\footnote{Note that for the range of
  momenta studied here the Gribov ambiguity is irrelevant as verified
  numerically in \cite{Sternbeck:2005tk}.} 
For the stopping criterion we chose 
$
  ~\max_x\, \Tr\left[(\nabla_{\mu} A_{x,\mu})(\nabla_{\mu}
    A_{x,\mu})^{\dagger}\right] < 10^{-13}.~
$
The fields $A_{x,\mu}\equiv A_{\mu}(x+\hat{\mu}/2)$ are the lattice gluon fields
given here in terms of gauge-fixed links $U_{x,\mu}$ by the mid-point definition
\begin{displaymath}
  A_{\mu}(x+\hat{\mu}/2) \;:=
  \frac{1}{2aig}(U_{x,\mu}-U^{\dagger}_{x,\mu}) -
  \frac{\identity}{6aig}\Tr(U_{x,\mu}-U^{\dagger}_{x,\mu})\,.
\end{displaymath}
This is accurate to order $O(a^2)$. On each such gauge-fixed
configuration the momentum-space gluon and ghost propagators were
measured. On the lattice, these are defined as the Monte Carlo averages
\begin{displaymath}
  D^{ab}_{\mu\nu}(k)=\left\langle
    \tilde{A}^a_{\mu}(k)\tilde{A}^b_{\nu}(-k)\right\rangle_{U} 
\quad\text{and}\quad
  G^{ab}(k) =
  \frac{1}{V}\left\langle\sum_{xy}\left(M^{-1}\right)^{ab}_{xy}\;e^{i
    k\cdot(x-y)}  
  \right\rangle_U
\end{displaymath}
where $\tilde{A}_{\mu}=\tilde{A}^a_{\mu}T^a$ are the
Fourier-transformed gluon fields, $M$ is the lattice Faddeev-Popov
operator in Landau gauge, and $ k\cdot x \equiv \sum_\mu 2\pi\, k_\mu
x_\mu/L_\mu $. For a definition of $M$ and details on its
inversion we refer to \cite{Sternbeck:2006rd,Sternbeck:2005tk} and
references therein. The corresponding bare dressing functions, $Z_L$ and
$J_L$, are then calculated by assuming a tensor structure for the
lattice propagators as given in \Eq{eq:gl_gh_tensor}, but with the
continuum momenta 
$q_{\mu}$ substituted by ~$p_{\mu}(k)=(2/a) \sin(\pi k_\mu/L_{\mu})$ where
the integers $k_{\mu}\in \{-L_\mu/2+1, \ldots \, L_\mu/2
\}$.\footnote{Of course, for both propagators the case 
  $k\equiv(k_1,k_2,k_3,k_4)=0$ has to be excluded.}
At tree level this tensor structure is exact for the Wilson gauge
action and the Faddeev-Popov operator we use (applying periodic
boundary conditions). Given the data for $Z_L$ and
$J_L$, we then estimate the coupling constant by the product
\begin{equation}
  \alpha_L(p^2) =
  \frac{g^2(a)}{4\pi}\, Z_L(p^2,a^2)\, J^2_L(p^2,a^2)
  \label{eq:alpha_s_def}
\end{equation}
where $g^2(a)=6/\beta$. The corresponding error is obtained from a
bootstrap analysis. Note that unlike other lattice investigations
where data for $Z$ and $J$ are usually separately renormalised, 
no renormalisation is done here.

\begin{table}
 \begin{minipage}{0.57\linewidth}
  \centering
   \begin{tabular}{cccc@{\quad}c}
    \hline\hline 
    $\beta$ & $\kappa$ & latt. & $a$ [fm] & \#conf. \\
    \hline
    5.25 & 0.13575 & $24^3\times48$ & 0.084 & 60 \\
    5.29 & 0.13590 & $24^3\times48$ & 0.080 & 55 \\
    5.40 & 0.13610 & $24^3\times48$ & 0.070 & 62 \\
    5.40 & 0.13640 & $32^3\times64$ & 0.068 & 57 \\
    5.40 & 0.13660 & $32^3\times64$ & 0.068 & 30 \\
    \hline
   \end{tabular}
 \end{minipage}
 \qquad
 \begin{minipage}{0.44\linewidth}
   \begin{tabular}{cccc@{\quad}c}
    \hline\hline 
    $\beta$ & latt. & $a$ [fm] & \#conf. \\
    \hline
    --   & --     & --    & -- \\
    6.20 & $32^4$ & 0.063 & 30 \\
    6.40 & $32^4$ & 0.048 & 50 \\
    6.60 & $32^4$ & 0.037 & 60 \\
    6.80 & $32^4$ & 0.029 & 46 \\
    \hline
   \end{tabular}
 \end{minipage}
    \caption{Number of $N_f=2$ (left) and $N_f=0$ (right)
     configurations used. To set the lattice spacings we
     use values for $(r_0/a)$ as provided by QCDSF
     \cite{Gockeler:2005rv,Zanotti:2007er} and assume $r_0=0.467$~fm.} 
\label{tab:stat}
\end{table}

\section{Preliminary results}
 
\begin{figure}[t]
  \centering
  \mbox{\includegraphics[height=5.6cm]{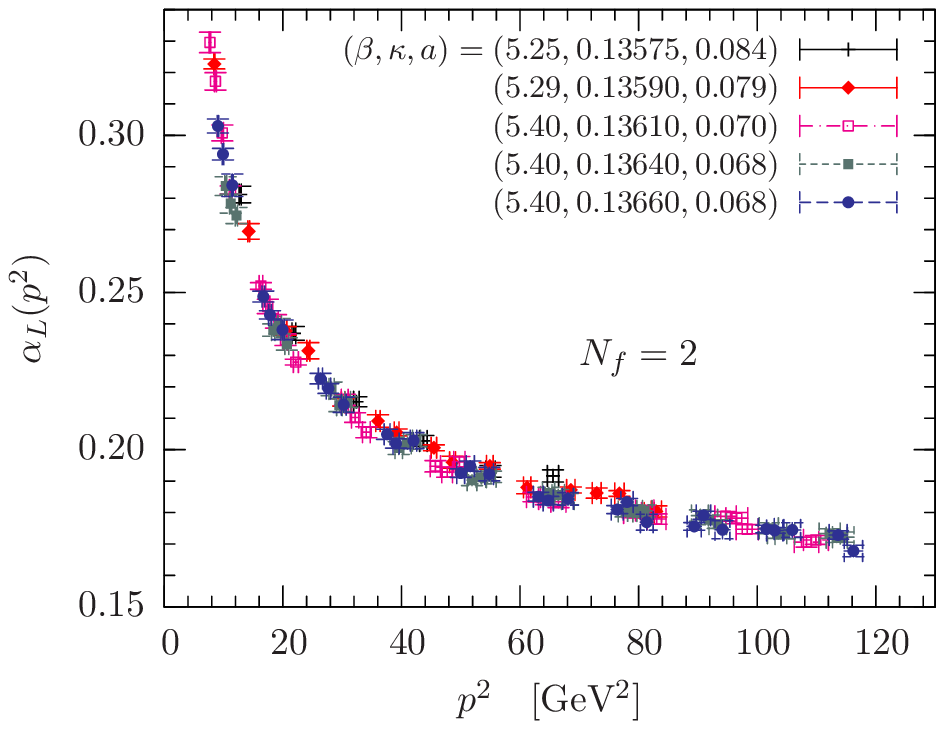}\quad
    \includegraphics[height=5.6cm]{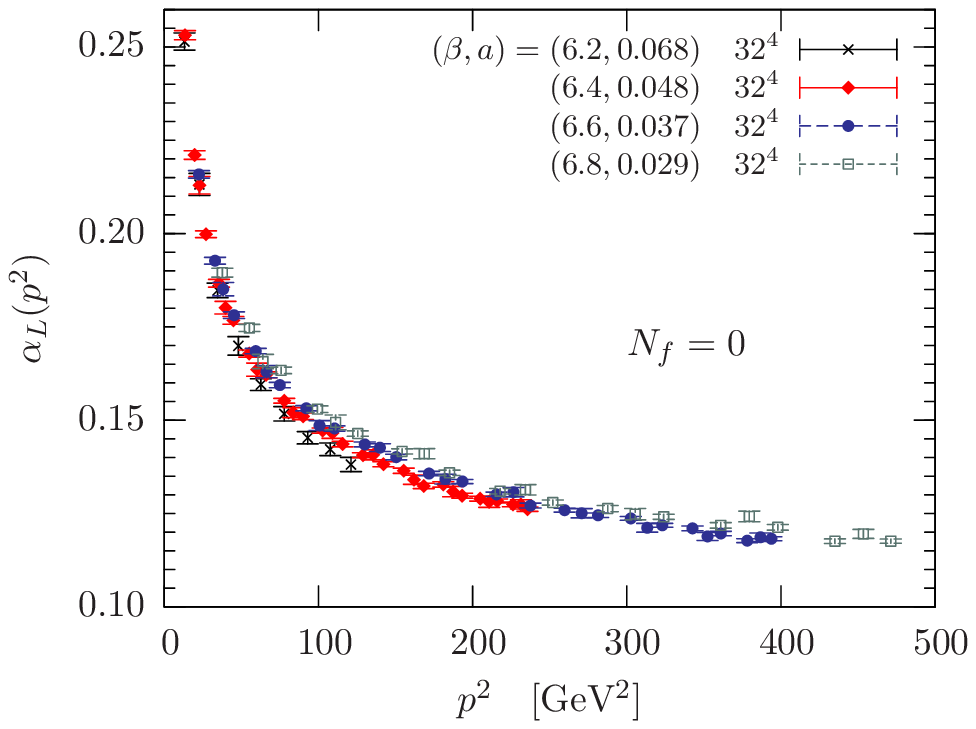}}
  \caption{$N_f=2$ (left) and $N_f=0$ (right) data for $\alpha_L(p^2)$
    at different $\beta$ (and $\kappa$). We set $r_0=0.467~\text{fm}$
    to assign physical units to $p^2$. Approximate values for the
    lattice spacing $a$ are given in fm.}
  \label{fig:alpha_qq_data}
\end{figure}

In \Fig{fig:alpha_qq_data} (left) we show our present $N_f=2$ data on
$\alpha_L$ as a function of momentum. One sees that the data for
different values of the input parameters $\beta$ and $\kappa$ form a
reasonably smooth curve, indicating that discretisation effects are
small for the input we employ. The only noticeable  exceptions are the
highest momentum values of each data set, where a slight deviation
becomes evident with decreasing $a$. 
As expected, the quark-mass dependence is small.

Discretisation effects are more evident for our present quenched data,
as shown in \Fig{fig:alpha_qq_data} (right). In particular for
$\beta\le 6.4$, which corresponds to even smaller lattice spacings
than we use for $N_f=2$, the data at larger momenta tends towards
larger values as $a$ decreases. Perhaps surprisingly, this effect is
in the opposite direction than that observed in the unquenched data. 

\begin{figure*}
  \begin{minipage}{0.48\linewidth}
  \centering
  \includegraphics[height=5.2cm]{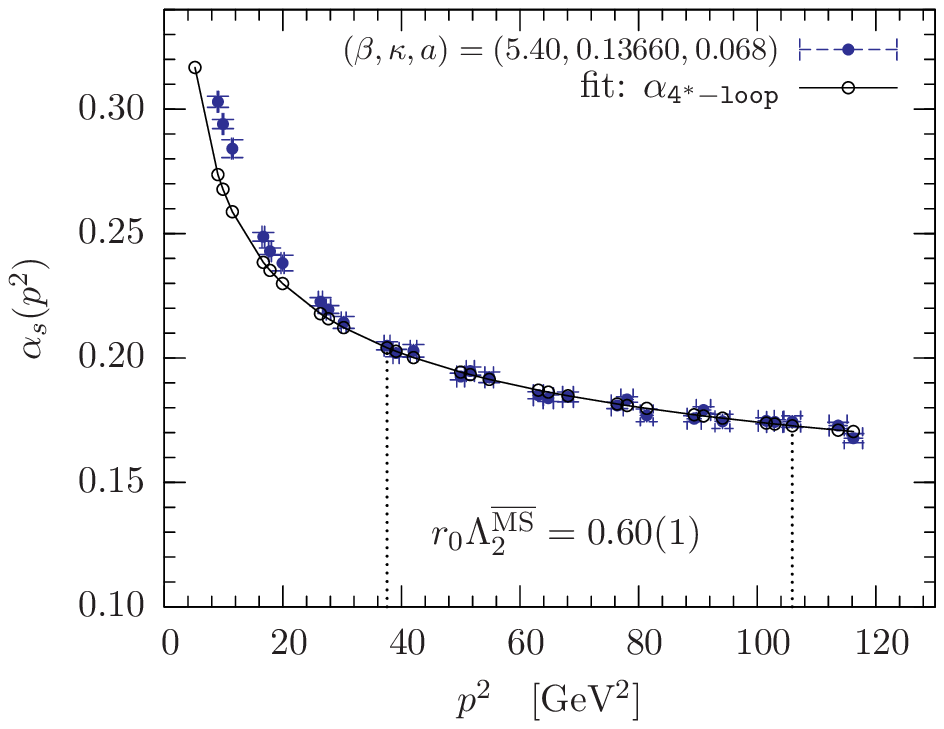}
  \caption{$N_f=2$ data for $\alpha_L(p^2)$ at
    $\beta=5.4$ and $\kappa=0.1366$. The
    line represents the best fit to the data as described in the text. The
    fit window and the value of $r_0\LambdaMS$, as a result of that
    fit, are given too.}
  \label{fig:alpha_qq_dyn_fit}
  \end{minipage}
  \quad
  \begin{minipage}{0.47\linewidth}
    \centering
    \includegraphics[height=5.2cm]{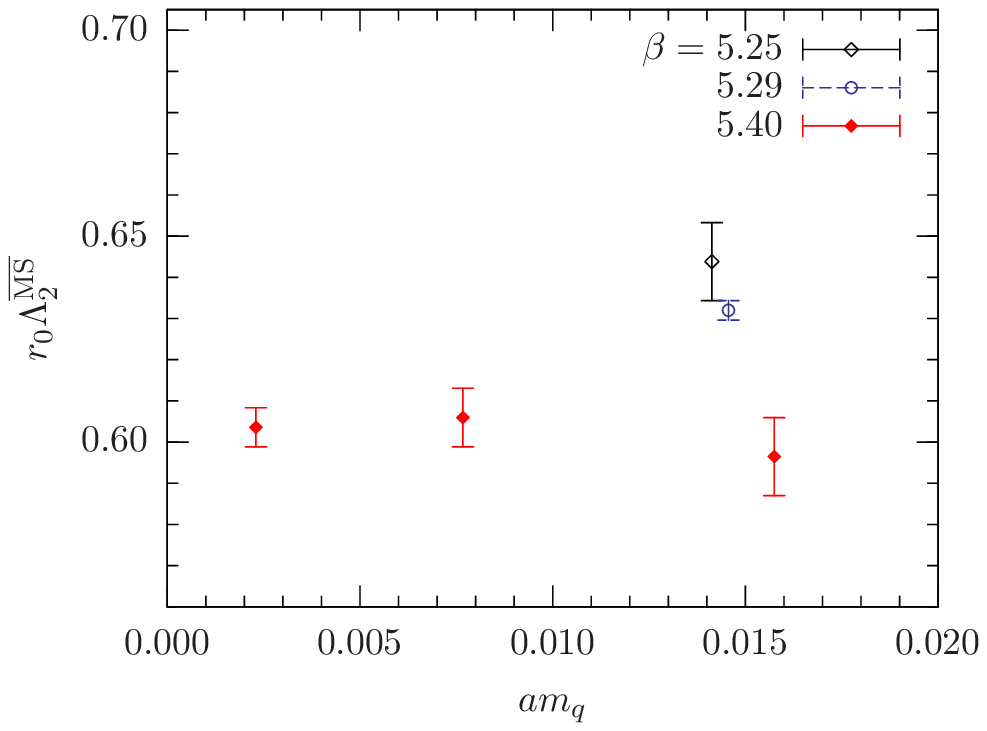}
    \caption{Fitted values for $r_0\LambdaMS_2$ as a function of 
      bare mass $am_q$ for different $\beta$. We use $\kappa_c$ values
      from \cite{Gockeler:2005rv,Zanotti:2007er} to set
        $am_q=(1/\kappa - 1/\kappa_c)/2$.\vspace{2.6ex}}
    \label{fig:r0Lambda2_mq}
  \end{minipage}
\end{figure*}

Since we use a tree-level improved definition of momentum (see above)
the data at larger momenta is expected to be less affected by
discretization errors than with the naive definition
\mbox{$a\tilde{p}_{\mu}=2\pi \, k_{\mu}/L_{\mu}$}. 
Nevertheless, if the data is fit to perturbative
QCD, e.g., as described below, the momenta $p_{\mu}$ considered must
satisfy  $
  ~\Lambda\sim 250~\textrm{MeV}\ll |p_{\mu}|\ll\pi/a~.
$
For the cylinder-cut data \cite{Leinweber:1998uu} used here, this translates
into an upper bound of $p^2 = p_\mu p_\mu  \ll  4\pi^2/a^2\sim
300~\text{GeV}^2$ with the lattice spacings for $N_f=2$ in
\Fig{fig:alpha_qq_dyn_fit}.  This leaves us a considerable range of
momenta to work with. 
We observe, however, that for both the quenched and the unquenched data,
strong discretisation effects appear at $a^2p^2\ge14$ (not shown), where
the data points (for fixed parameters) start bending downwards as $a^2p^2$ is
further increased.  We therefore restrict the present analysis to momenta
$ a^2p^2<14$ where no such effect occurs.

Given the data for different parameters, we consider, for each
data set, a range of fitting windows within which the data
is fit to a perturbative
expansion of our running coupling. Since, from perturbative QCD, the running of
$\alphaMOM$ is known up to three loops for the ghost-gluon vertex
\cite{Chetyrkin:2000fd}, we could use the
truncated $\alphaMOM$ beta-function at 3-loop order from
Ref.~\cite{Chetyrkin:2000fd} to describe the running of our
data. However, the coefficients $c_1$ and $c_2$ in the expansion of
$\alphaMOM$ in terms of $\alphaMS$, i.e.\ 
\begin{equation}
  \label{eq:alphaMOM_expansion}
  \alphaMOM = \alphaMS\Big\{1+c_1 \alphaMS + c_2\big[\alphaMS\,\big]^2 +
  c_3\big[\alphaMS\,\big]^3 + \ldots\Big\}\;,
\end{equation}
which are known, as a function of $N_f$ \cite{Chetyrkin:2000fd}, 
\footnote{For the ghost-gluon vertex those coefficients are roughly
  $c_1(0)\approx4.23/\pi$ and $c_2(0)\approx 36/\pi^2$ for the
  zero-flavour and $c_1(2)\approx3.67/\pi$ and $c_2(2)\approx
  26/\pi^2$ for the two-flavour case \cite{Chetyrkin:2000fd}.
}
are all positive and of $O(1)$ for the cases considered here, causing
$\alphaMOM$ to run more rapidly with scale than does $\alphaMS$.
The result is that, as one lowers the scale,
the truncated running at a given order becomes problematic at higher
scales for the $\MOM$ coupling than it does for the $\MSbar$ coupling. 
In order to opt for the safest version of our analysis from the outset, 
we thus perform the running using the intermediate $\alphaMS$ running 
at four loops, with matching from $\alphaMOM$ to $\alphaMS$ at the start, 
and then re-matching back to $\alphaMOM$ at the end. 

In practice, this means that for our fits we scan over a fine-grid
interval of $\alphaMS $ values at an arbitrary reference scale
$\mu$, running each such value to all the momenta $p^2$ considered with the
truncated 4-loop running%
\footnote{To be specific, the running is performed using the exact
  analytic (implicit) solution of \Eq{eq:RGE} corresponding to the
  four-loop truncated beta function.} relevant to $N_f$.
At each $p^2$ the 4-loop value of $\alphaMS(p^2)$ is then related to the
corresponding $\MOM$ value $\alphaMOM(p^2)$ using
the relation between the $\MOM$ and $\MSbar$ couplings given in
\Eq{eq:alphaMOM_expansion}. For this we use the known coefficients
$c_1$ and $c_2$ and set \mbox{$c_3=c_4=\ldots=0$}, an approximation which
appears to become reliable for sufficiently large $p^2$ in our data sets.
A $\chi^2$-minimisation of the scaled deviations of $\alphaMOM(p^2)$
from the $\alpha_L(p^2)$ data then determines the optimised
  $\alphaMS(\mu^2)$, from which the 
optimised $\LambdaMS$ value is obtained using the conventional
definition given, e.g., in Ref.~\cite{Chetyrkin:1997sg}. 
We perform such a fit separately on the data for each ($\beta$,
$\kappa$, $N_f$). An example of the resulting fit quality, together
with the corresponding value for $r_0\LambdaMS$, is shown for the 
$(\beta,\kappa, N_f)=(5.4,0.1366,2)$ set in
\Fig{fig:alpha_qq_dyn_fit}. Obviously, the data there is well
described by the fit. The result shown there, $r_0\LambdaMS=0.60(1)$,
is already in the ballpark of what is expected based on the results
of, e.g., Refs.~\cite{Gockeler:2005rv,DellaMorte:2004bc}. 
In \Fig{fig:r0Lambda2_mq} all our present (as-yet-preliminary) values
of $r_0\LambdaMS$ for the $N_f=2$ data sets are collected. Note that
we have yet to fully investigate and quantify, and hence have not
included in \Fig{fig:r0Lambda2_mq}, uncertainties due to (i) truncated
running, (ii) the impact of possible higher order, non-zero $c_3, c_4,
\ldots$ terms in the relation between the $\MSbar$ and $\MOM$ 
couplings, and (iii) the statistical uncertainty in the fit associated
with that in the data. In particular, this will then allow us to 
extrapolate our $r_0\LambdaMS$ values for the different parameters to the
appropriate limits ($a\to 0,\ \kappa \to \kappa_c$) with realistic
 error estimates.

\section{Conclusions}

We have reported on first steps towards a determination of
$\LambdaMS$ in terms of lattice MC simulations of gluodynamics within 
the Landau gauge. Our method is based on the 
ghost-gluon vertex which in this particular gauge provides a nonperturbative
running coupling in a MOM scheme defined solely in terms of
the gluon and ghost dressing functions. Both these dressing functions
can be calculated in terms of lattice MC simulations with good accuracy.

Although our results are still preliminary, fits of our data to
corresponding perturbative expressions of our running coupling result
in values of $\LambdaMS_2$ in the range expected. This
suggests that a full estimate of the QCD parameter $\Lambda$ within
this framework is worth pursuing. 
Different systematic effects still have to be investigated before final
conclusions can be drawn. In addition to those already noted above,
lattice discretisation errors, in particular for $N_f=0$, need further study.
In the light of the results shown in \Fig{fig:alpha_qq_dyn_fit}, however,
the approach appears quite promising.
\clearpage

{\small
We thank the QCDSF collaboration for providing us their $N_f=2$ gauge
configurations. This research was supported by the Australian Research
Council and by the Deutsche Forschungsgemeinschaft through the
Research Group {\it Lattice Hadron Phenomenology} (FOR 465). K.~M.\
acknowledges the ongoing support of the Natural Scienes and
Engineering Council of Canada. Generous grants of time on the IBM
pSeries 690 at HLRN (Germany) are acknowledged. 
}


\providecommand{\href}[2]{#2}\begingroup\raggedright
\endgroup

\end{document}